\documentclass[10pt]{iopart} 
\usepackage{graphicx}
\begin{document}

\title[The Fermi-Pasta-Ulam numerical experiment]
{The Fermi-Pasta-Ulam ``numerical experiment'': history and
pedagogical perspectives}

\author{Thierry
  Dauxois\dag \footnote[3]{To whom correspondence should be addressed
(Thierry.Dauxois@ens-lyon.fr)}, Michel Peyrard\dag\ and Stefano
  Ruffo\dag\ddag}

\address{\dag\ Laboratoire de Physique, UMR-CNRS 5672, ENS Lyon,
46  All\'{e}e d'Italie, 69364 Lyon c\'{e}dex 07, France}

\address{\ddag\ Dipartimento di Energetica, ``S. Stecco'' and CSDC,
  Universit{\`a} di
  Firenze, and INFN, via S. Marta, 3, 50139 Firenze, Italy}

\begin{abstract}
The Fermi-Pasta-Ulam (FPU) pioneering  numerical experiment played a major role in the
history of computer simulation because it introduced this concept for
the first time. Moreover, it raised a puzzling question which was
answered more than 10 years later. After an introduction to this
problem, we briefly review its history and then suggest some simple
numerical experiments, with a provided Matlab$^\copyright$ code, to study various
aspects of the ``FPU'' problem.

\end{abstract}




\section{The physical question}

The ``FPU problem'', as it is known presently, bears the name of
the three scientists\footnote{Mary Tsingou, who took part in the
numerical study is not an author of the report, but her
contribution is however recognised by two lines of
acknowledgments!} who wanted to study the thermalization process of a
solid~\cite{FPU}. As revealed by S. Ulam later~\cite{Ulam}, these
authors were looking for a theoretical physics problem suitable
for an investigation with one of the very first computers, the
``Maniac''. Their work, published in 1955 in a classified Los
Alamos National Laboratory report, had been completed shortly before the
death of Enrico Fermi in 1954. Ulam said later that Fermi did not
suspect the importance of this discovery and considered this work
as ``minor''. He nevertheless intended to talk about it at the
conference of the American Mathematical Society, where he had been
invited just before he became seriously ill.

Fermi, Pasta and Ulam decided to study how a crystal evolves towards
thermal equilibrium by simulating a chain of
particles of unitary mass, linked by a
quadratic interaction potential, but also by a weak nonlinear
interaction. One of the one-dimensional systems they considered
is described by the
Hamiltonian
\begin{equation}\label{hamilFPU}
  H=\sum_{i=0}^{N}\frac{1}{2}p_i^2
+\sum_{i=0}^{N}\frac{1}{2}(u_{i+1}-u_{i})^2
  +\frac{\alpha}{3}\sum_{i=0}^{N}(u_{i+1}-u_{i})^3\quad,
\end{equation}
where $u_i$ is the displacement along the chain of atom $i$ with respect to
its equilibrium position, and  $p_i$ its
momentum. The stiffness of the harmonic spring
and the lattice spacing have been set to one,
without losing generality.
The coefficient  $\alpha\ll 1$ measures the strength of the
nonlinear contribution to the interaction potential. The two ends of
the chain were assumed to be fixed, i.e. $u_0=u_{N+1}=0$.

The common approach in physics is to think in terms of ``normal
modes'', related to the displacements through
$A_k=\sqrt{{2}/({N+1}})\sum_{i=1}^{N}u_i\sin\left(ik\pi/{N+1}\right)$
with the frequencies
$\omega_k^2=4\sin^2\left({k\pi}/{2(N+1)}\right)$.
They provide a basis to rewrite the Hamiltonian~(\ref{hamilFPU}) as
\begin{equation}\label{hamilFPUfourier}
  H=\frac{1}{2}\sum_{k=1}^N \left(\dot
  A_k^2+\omega_k^2A_k^2\right)+\frac{\alpha}{3}\sum_{k,\ell,m=1}^N c_{k\ell m}A_kA_\ell
  A_m\omega_k\omega_\ell
  \omega_m\quad,
\end{equation}
where the coefficients $c_{k\ell m}$ are given for example in Ref.~\cite{poggi}.
The last term, generated by the nonlinear contribution to the potential,
leads to a coupling between the modes, and scales as $N^{3/2}$.

\begin{figure}[h!]
\centerline{\includegraphics[height=6.0cm]{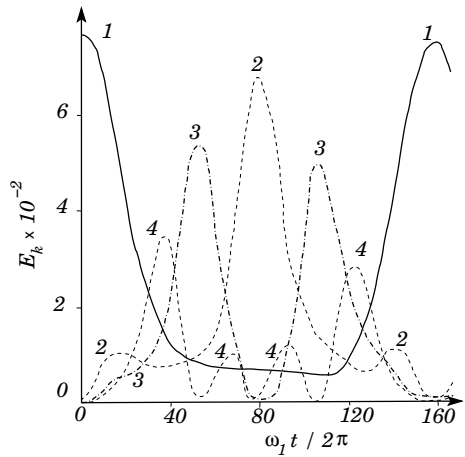}}
\caption{\small FPU recurrence: the plot shows the time evolution
of the kinetic and potential energy $E_k=\frac{1}{2}\left(\dot
  A_k^2+\omega_k^2A_k^2\right)$ of each of the four lowest
  modes. Initially, only mode~1 was excited (from
  reference~\cite{FPU}). }
 \label{fig:FPU}
\end{figure}

FPU thought that, due to this term, the energy introduced into a single
mode, mode $k= 1$ in their simulation, should slowly drift to the
other modes, until the equipartition of energy predicted by
statistical physics is reached.
The beginning of the calculation indeed suggested that this would be the
case. Modes 2, 3, \ldots, were successively excited.
However, by accident~\cite{Metropolis},
one day, they let the program run long after the steady state had
been reached. When they realized their oversight and came back to
the room, they noticed that the system, after remaining in a
steady state for a while, had then departed from it. To their
great surprise, after 157 periods of the mode $k= 1$, almost all the
energy (all but 3$\%$) was back to the lowest frequency mode, as
shown in figure~\ref{fig:FPU}.

The initial state seems to be almost perfectly recovered after this
recurrence period. Further calculations, performed later with
faster computers, showed that the
same phenomenon repeats many times, and that a
``super-recurrence'' period exists, after which the initial state is
recovered with a much higher accuracy. Thus, contrary to the expectations
of the authors, the drift of the energy of the initially excited mode~1
towards the modes with a higher wave number does not occur.
This highly remarkable result, known as the FPU paradox, shows that
nonlinearity is not enough to guarantee equipartition of energy.

\section{Fermi, Pasta, Ulam: the characters.}

Born in Rome, {\sc Enrico Fermi} (1901-1954)
has been one of the brightest physicists of the twentieth century,
who made major experimental and theoretical discoveries. His name
is famous for his contributions to statistical physics, elementary
particle physics, the control of nuclear energy. It is thanks to
his trip to receive the Nobel prize in Sweden in 1938 that he
left fascist Italy, and emigrated to the
United States where he studied atomic fission and set up the first
controlled chain reaction in Chicago in 1942. He was naturally
called to be one of the leaders of the Manhattan project for the
development of nuclear energy and the atomic bomb. We are less
familiar with his work on various nonlinear problems at the end of
his life, before his premature death from stomach cancer.

{\sc John R. Pasta} (1918-1981) did not have such a bright
career. In spite of his interest in physics he had to leave the
New York City College during the great depression. He was first a
real estate agent and then a police officer in New York from 1941
to 1942. Then he has been recruited as radar and cryptography
specialist by the American army during the second world war.  His
earnings as a GI allowed him to return to the University, where he
got a PhD in theoretical physics in 1951. He immediately started
to work in Los Alamos, on the MANIAC computer which was under
construction and tests. His career went developed as a computer expert
for the Atomic Energy Commission, where he extensively developed
the Mathematics and Computer Division. Then John R. Pasta became a
physics professor and, in 1964, dean of the computer science
department of the university of Illinois, where he focused on the
use of computers to solve applied problems in physics and
mathematics.

Born in Poland, {\sc Stanislaw M. Ulam} (1909-1984) quickly became
interested in mathematics and obtained a PhD in 1933 under the
supervision of  Banach (1892-1945). Following a first invitation
by Von Neumann to the prestigious Princeton Institute for Advanced
Study in 1935, he visited the Unites States several times before
leaving Poland definitely before the start of the second world
war. He became an American citizen in 1943, and was invited by von
Neumann himself to join the Los Alamos team to prepare the atomic
bomb. Among other things he solved the problem of the initiation
of fusion in the hydrogen bomb. In collaboration with  N. C.
Metropolis he invented the Monte-Carlo method to solve problems
requiring a statistical sampling. He stayed in New Mexico until
1965 before his appointment as a mathematics professor at
Colorado University.

The FPU numerical experiment has been performed on the {MANIAC}
(Mathematical Analyser, Numerical Integrator And Computer) built in
1952 for the Manhattan project, which has been used in the development
of  Mike, the first hydrogen bomb. Richard Feynman (1918-1987) and
Nicholas C. Metropolis (1915-1999), exasperated by the slow and
noisy mechanical calculators which were nevertheless necessary to the
design of the bombs, promoted its construction. The name of the
computer, MANIAC, had been chosen by the project director,
N. Metropolis, who hoped that it would stop the raising fashion of
naming computers by acronyms. The effect has been
exactly the opposite, although,
according to Metropolis~\cite{Metropolis}, George Gamow
(1904-1968) was instrumental in rendering this and other computer
names ridiculous when he dubbed the Maniac ``Metropolis And von
Neumann Install Awful Computer''. The
Maniac was able to perform about  $10^4$ operations per second, which
must be compared to the  $10^8$ operations per second of any personal
computer today.
It is Fermi who had the genius to propose that computers could be used
to study a problem or test a physical idea by simulation, instead of
simply performing standard calculus. He proposed to check the
prediction of statistical physics on this system now called FPU.

The discovery that resulted from this study has not only been at the
origin of the soliton concept and of many features of chaotic
phenomena, as discussed in the following section,
but it also introduced the concept of {\em numerical
experiment.} This had far reaching consequences, leading to a
complete revolution in the investigation of physical phenomena. The
computer is no more used only to perform a calculation that cannot be
done by pencil-and-napkin, but to check a theoretical conjecture that
cannot be proven analytically, or even to provide the theorist with
``experimental" results that wait for a mathematical proof: a source
of problems, like in a ``true laboratory experiment". Of course, the
{\em numerical experiment} does not possess all the complexity of a
true physical experiment, because the reality which is represented is
highly virtual. However, today numerical simulations of
condensed matter systems, one reaches such a level of confidence that
sometimes it has happened that a laboratory experiment has been
questioned because a numerical experiment had given contrary
indications~\cite{jensen}. Today, {\em computational physics} is an
established discipline and it is considered as sort of separated from
both theoretical and experimental physics. Students are
currently trained in computational physics as in other disciplines,
and specialized journals publish the results of the research in this
field.  This big epistemological
and sociological change began with the FPU experiment.

\section{Epilogue: The solution of the FPU problem}

Pursuing the solution of the FPU paradox, two main different lines of
thought were followed. On one side, people like J. Ford~\cite{Ford1}
 focused on the Fourier mode
dynamics, looking for non-resonance conditions that could explain the
inefficient energy transfer.  No convincing explanation was found
before the news spread of a theorem announced by
Kolmogorov~\cite{Kolmogorov}, and then proven by Moser~\cite{Moser}
and Arnold~\cite{Arnold} (KAM theorem), which states that most orbits
of a slightly perturbed integrable Hamiltonian systems remain
quasi-periodic.  Although {\em qualitative} explanations of the FPU
recurrence were obtained following this path no {\em quantitative} explanation has been obtained,
to our knowledge.
On the contrary, a remarkable result was obtained by Izrailev and
Chirikov~\cite{Chirikov1}, following KAM theory.  If the perturbation
is strong enough that nonlinear resonances ``superpose", the FPU
recurrence is destroyed and one obtains a fast convergence to thermal
equilibrium.  This prediction was tested numerically and is nowadays
at the basis of the phenomenon called ``strong stochasticity
threshold"~\cite{Livi85}.

The other line of thought, which went towards the so-called {\em
continuum limit}, led to the solution of the FPU paradox by Zabusky
and Kruskal~\cite{Zabuskykruskal} in terms of the dynamics of {\em
solitons}~\cite{DauxoisPeyrard}.  Starting from the FPU equations of motion, derived from
Hamiltonian~(\ref{hamilFPU}),
\begin{equation}\label{eqmotionFPU}
  \ddot
  u_i=(u_{i+1}+u_{i-1}-2u_{i})+\alpha
\left[(u_{i+1}-u_{i})^2-(u_{i}-u_{i-1})^2\right]\quad,
\end{equation}
and restricting the investigation to long wavelength modes,
they could derive
the Korteweg-de Vries equation~\cite{KdV}
\begin{equation}\label{equationFPUKdV}
  w_\tau+\frac{1}{24}w_{\xi\xi\xi}+\alpha \,ww_\xi=0 \quad ,
\end{equation}
where the field $w$ is a conveniently
rescaled spatial derivative of the displacement field $u_i$,
$\tau$ is a rescaled time and $w_x=\partial w/\partial x$.

The sinusoidal initial
condition develops sharp fronts, and then  breaks into a series of
pulses, which are solitons. They preserve their
shapes and velocities and, during their motion in the finite system with
periodic boundary conditions\footnote{The calculation performed by
Zabusky and Kruskal used periodic boundary conditions, while the
FPU calculation used fixed boundaries. This does not change the
analysis because either the solitons pass through the boundaries
and enter on the opposite side, or they are totally reflected},
from time to time, they come back to the positions that they had
initially, restoring the initial condition. This is why FPU recurrence
is observed.

It is important to emphasize that the FPU paradox would probably not
have been a mystery for more than 10 years, if, before Zabusky and
Kruskal, somebody had the idea to look carefully at the dynamics of the
nonlinear lattice as a function of the space coordinate (see the suggested
numerical experiment in Sec.~\ref{pedagoperpect}). But
physicists were so used to analyse linearised problems, which
naturally leads to a description in the Fourier space, and only add
nonlinearity afterwards as a coupling between the modes, that the
observation of the solitons emerging from the sinusoidal initial
condition had not been done.

\section{The remake: FPU and the Japanese School}

The history of the FPU problem is quite rich, as documented
in the report by Ford~\cite{Ford92} and in the recent book
by Weissert~\cite{Weissert}. We will not discuss it in this
paper and we just recall that some of the contributions
to the Chaos issue~\cite{Chaos}, printed this year to celebrate
the anniversary of the FPU experiment, are devoted to a brief
historical reconstruction.

Because of its relevance for the determination of a quantitative
formula for the recurrence time, we would like to discuss briefly the
contribution of the Japanese school, which has been very active
in this field since the 60's.

The original FPU report~\cite{FPU}, not published, was known by very
few people, most of them within
America. This is why, in Japan, Nobuhiko Saito  (1919-) together with his PhD
student Hajime Hirooka  (1939-) were developing closely
related research works in complete ignorance of what had been
done in Los Alamos ten years earlier.

As explained by the title of Hirooka's PhD Thesis, ``The approach
to thermal equilibrium in a nonlinear lattice'', the motivation
was to explore the mechanism of ergodicity. As the mechanics of
collisions in a gas was too complicated, they thought that an
anharmonic lattice would be a more appropriate system; it appeared
finally that its dynamics was difficult to solve. This is why, in 1964
they started to perform numerical simulations on a small
computer provided by NEC: it took all the night to compute the
evolution for only 5 lattice sites during several periods of the
lowest mode! In 1965, they switched to the new Supercomputer
provided by Hitachi Co. to Tokyo University, installed under a
national plan for all Japanese universities and research
institutes.

Saito and Hirooka considered a one-dimensional anharmonic lattice
with quadratic and quartic potentials between neighboring
particles, and with both ends fixed. The initial excitation was
slightly different since all particles were at rest, whereas a
constant force was applied to the first particle. They also
prepared a similar system, with only harmonic potentials, and
analytically calculated several quantities of interest, in
particular the long-time averages of the squares of the velocities
of the particles $\langle \dot u_i^2 \rangle$ and the correlations
$\langle \dot u_i \dot u_{i+1} \rangle $
 of the velocities of neighboring particles.
The long-time averages of  $\langle \dot u_i^2 \rangle$  were the
same for all particles, but the correlation functions did not
vanish in the long term, and the Maxwellian
distribution of velocities was not observed, contrary
to their expectations.

It is interesting to emphasize that they were hesitating to
publish these results, because they were not familiar with the
computer calculation and were afraid of having introduced some
errors. However, they found numerical experiments
similar to the FPU original one in several papers on nonlinear
oscillation, in particular, those of Ford~\cite{Ford1}.
These papers also convinced Morikazu Toda to study solitons and to
introduce the lattice with exponential interactions, nowadays
called the Toda lattice~\cite{Todaexplattice}.

Soon after, they also learned that the original FPU
report~\cite{FPU} was reproduced in the Collected Papers of Fermi.
Very excited, Saito and Toda read the paper at the library of
Tokyo University of Education (now Tsukuba University). As they
had found results similar to those of Fermi, Pasta and
Ulam, they finally decided to publish their
calculations~\cite{Saito1}. Afterwards, they considered the
simpler FPU initial condition and found the induction
phenomenon and the occurrence of the random character of lattice
vibrations.

Finally, they discovered the Zabusky-Kruskal seminal
paper~\cite{Zabuskykruskal} soon after the publication of their
papers in 1967. Norman Zabusky went to Kyoto in 1968 at the
International conference of Statistical Mechanics, where he showed
his movies and in particular the  KdV cinema, which ``has the
power to communicate unbiased information in a credible manner,
much beyond the power of words, graphs and
equations''.

At that time, M. Toda gave the first
analytical estimate of the recurrence time, based on the exact
solutions of the exponential lattice, nowadays called the Toda
lattice. Introducing  the period $T_1=2N$ of the first
mode and  the amplitude $a$ of the initial sine wave, he got
\begin{equation}\label{recurence}
  T_R= \frac{3}{\pi^{3/2}\sqrt{2}}\frac{N^{3/2}}{\sqrt{a\alpha}}\
  T_1\simeq 0.76\ \frac{N^{5/2}}{\sqrt{a\alpha}}\quad,
\end{equation}
which compares well with the first empirical estimate
due to Zabusky.
 Toda's result was a little
larger than Zabusky's formula: the prefactor is 0.31 instead of
0.38 in equation~(\ref{recurence}). The discrepancy is due to the
fact that, when solitons pass through each other, they accelerate
because of the compression of the lattice. See Ref.~\cite{Toda}
for a more complete discussion and Ref.~\cite{Goedde} for a recent
check of formula (\ref{recurence}).

\section{Pedagogical perspectives}
\label{pedagoperpect}

It is quite easy, with contemporary computational tools, to reproduce
the original FPU experiment.  An example is the simple MATLAB$^\copyright$ code
reported in the Appendix, which solves the equations for the dynamics
of the FPU model and computes the normal modes $A_k$.
We suggest the student to use it
to reproduce different aspects of the FPU problem
\begin{itemize}

\item Try to reproduce Fig.~\ref{fig:FPU} using the original FPU
initial condition given in the code.  Be careful in the choice of the
amplitude $a$ because the cubic lattice is unstable at large energies.

\item By careful inspection of the difference field
$(u_{i+1}-u_{i})$, it is possible to detect the presence of
propagating structures, which are indeed the solitons of the
Zabusky-Kruskal analysis.

\item Between  $a=1$ and $a=10$, the FPU model with $N=32$ undergoes
relaxation to equilibrium~\cite{Chirikov1}. Check the absence of FPU recurrences
and the establishment of energy equipartition among the normal modes.

\item Since the initial sine wave form solitons~\cite{Zabuskykruskal}, it is quite natural
to begin with a lattice soliton initial
condition. The exact kink-like soliton solution for the Toda lattice~\cite{Toda} is
\begin{equation}u_i(t)=\pm\frac{1}{2\alpha}\log\frac{1+\exp[2k(i-1-i_0)\pm t\sinh k]}
{1+\exp[2k(i-i_0)\pm t\sinh k]},
\end{equation}
where $k$ is the inverse width. This solution is only approximate
for the FPU lattice, and due to the fixed boundary conditions, one
must put a kink (+) and an antikink (-) solutions, centered around
2 different sites of the lattice. Our advice is to first choose a
large lattice (e.g. $N=128$). The time evolution of the derivative
field $u_{i+1}-u_i$ shows the propagation of coherent structures,
their mutual interactions and reflection from the boundaries.

\item Try to reproduce the relevant scalings of formula
(\ref{recurence}), for instance the $N^{2.5}$ dependence on the number
of oscillators.

\item Finally, try the Zabusky-Deem~\cite{zabuskydeem} initial condition $u_i=a \sin[i\pi N/(N+1)]$,
which corresponds to the highest frequency mode. Similar oscillations of mode energies as the one
of long wavelengths are observed. Try to characterize these recurrences.

\end{itemize}

 The scaling of formula~(\ref{recurence}) is quite different from the
one of the Poincar\'e recurrence time~\cite{Poincare}, which can be
shown to increase exponentially with $N$ in the harmonic
limit~\cite{Hemmer}.  No striking change of this scaling should occur
for a small nonlinearity, e.g.  in the FPU recurrence region, although
nothing is rigorously known for this case.  The interested student can
find the statement of Poincar\'e recurrence theorem in most books of
statistical mechanics. However, in simple words, the theorem says that
whenever a dynamical system preserves phase-space volume in its time
evolution (and this is the case of Hamiltonian systems, including the
FPU oscillator chain), ``almost all" trajectories return arbitrarily
close to their initial position, and they do this an infinite number
of times.  The exponential scaling with $N$ of the Poincar\'e
recurrence time with respect to the algebraic one of Toda's formula
proves that the FPU recurrence is not a manifestation of this
phenomenon. It's remarkable that, although certainly Fermi and Ulam
did know Poincar\'e theorem, they did not quote it at all in their
paper as a possible explanation of the observed
recurrence\footnote{This fallacious possibility has been considered by
others (who perhaps did not know the theorem as well as Fermi and
Ulam) when the news about the FPU experiment spread around.}.

Another pedagogical aspect, which is worth discussing in
connection with the FPU experiment, is the relation between {\it
continuum} and {\it discrete}. Analytical ``human'' approaches to
the problem consider continuous variables, while with the
computers, it was natural and necessary to use discretized
variables; the human and computer way of dealing with physical
problems were therefore for the first time contrasted. The only
discreteness appearing in the FPU model is the spatial one: the
oscillators are put on a lattice. Time flows continuously and the
displacement field variable $u_i$ is also continuous. Models of
space-time evolution have been considered where time is discrete,
while the lattice variable is still continuous: so-called {\it
coupled map lattices}~\cite{ChaosCML}. Hamiltonian versions of
this type of models exist, in the sense that phase-space volume is
conserved by the dynamical evolution, although the total energy is
not.  The extreme case of discreteness is the one of {\it cellular
automata}~\cite{Wolfram}, for which also the lattice variable
takes a discrete set of values.  Depending on the physical system
at hand, one model or the other has been considered as
appropriate, and several examples exist where even the extreme
discrete case of cellular automata provides a good representation
of the physical process under analysis.

In this respect, quite advanced still open questions related to the
FPU recurrence are: Is the recurrence observed also in ``Hamiltonian"
coupled map lattices and in cellular automata? Is it related to the
propagation of coherent structures which have some resemblance to
solitons? How does the recurrence time scale with system size? These
are open research issues, on which perhaps a curious PhD student might
want to challenge his own computational skill, taking the
opportunity to make his own first ``numerical experiment", a privilege
once reserved only to a few people isolated in a deserted region of
New Mexico.

\ack We  thank R. Botet and E. Trizac for the opportunity they
gave us to contribute to this \EJP issue on ``Teaching Physics
with computers''. We also thank G. Battimelli, H. Hirooka, M. D.
Kruskal, N. Saito and N. J. Zabusky for useful exchanges of
information. We thank R. Khomeriki for help in coding with Matlab.
S.R. thanks ENS Lyon for hospitality and CNRS for financial
support. This work is part of the contract COFIN03 of the Italian
MIUR {\it Order and chaos in nonlinear extended systems}.

\section*{Appendix: MATLAB$^\copyright$ Code}

Codes are available at the website:\\ {\tt
http://perso.ens-lyon.fr/Thierry.Dauxois/fpu.html}

\subsection*{Main Program}

{\obeylines \tt\parindent=0pt
N=32;          \hfill \% Number of particles must be a power of 2
alpha=0.25;     \hfill \% Nonlinear parameter
TMAX=10000;  DT=20;  \hfill \% tmax and Delta t
tspan=[0:DT:TMAX];
options=odeset('Reltol',1e-4,'OutputFcn','odeplot','OutputSel',[1,2,N]);
\% Test different tolerances, changing Reltol
for I=1:N,

a=1; b(I)=a*sin(pi*I/(N+1)); b(I+N)=0;  \hfill\% FPU initial condition

\%a=1; b(I)=a*sin(pi*N*I/(N+1)); b(I+N)=0;\hfill\% Zabusky-Deem init. cond.

\%k=0.8; sk=(sinh(k))\verb!^!2; ek=exp(-k); i1=I-N/4; i2=i1-N/2;\hfill \%Solitons
\%b(I)=-0.5/alpha*log((1+exp(2*k*(i1-1)))/(1+exp(2*k*i1))); 

\%b(I)=b(I)+0.5/alpha*log((1+exp(2*k*(i2-1)))/(1+exp(2*k*i2))); 

\%b(I+N)= sk*ek/alpha/cosh(k*i1)/(exp(-k*i1)+exp(k*i1)*exp(-2*k));
\%b(I+N)=b(I+N)-sk*ek/alpha/cosh(k*i2)/(exp(-k*i2)+exp(k*i2)*exp(-2*k));

omegak2(I)=4*(sin(pi*I/2/N))\verb!^!2;  \hfill\% Mode Frequencies
end
[T,Y]=ode45('fpu1',tspan,b',options,N); \hfill\% Time integration
for IT=1:(TMAX/DT),
  TIME(IT)=IT*DT*sqrt(omegak2(1))/2/pi; \hfill\% Time iteration loop
  YX(IT,1:N+1)=[0 Y(IT,1:N   )];  YV(IT,1:N+1)=[0 Y(IT,N+1:2*N   )];
  sXF(IT,:)=imag(fft([YX(IT,1:N+1) 0  -YX(IT,N+1:-1:2)]))/sqrt(2*(N+1));
  sVF(IT,:)=imag(fft([YV(IT,1:N+1) 0  -YV(IT,N+1:-1:2)]))/sqrt(2*(N+1));
  Energ(IT,1:N)=(omegak2(1:N).*(sXF(IT,2:N+1).\verb!^!2)+sVF(IT,2:N+1).\verb!^!2)/2;
  for J=2:N-1,  \hfill\% Space loop
    DifY(IT,J)=Y(IT,J+1)-Y(IT,J);
  end
end

plot(TIME,Energ(:,1),TIME,Energ(:,2),TIME,Energ(:,3),TIME,Energ(:,4));

surf(DifY); \hfill\% Space derivative field to show the soliton dynamics }
\subsection*{fpu1 function}

{\obeylines \tt\parindent=0pt
function dy=fpu1(t,y);
N=32;alpha=0.25;
D(N+1)=y(2)  -2*y(1)+alpha*((y(2)-y(1))\verb!^!2-y(1)\verb!^!2);D(1)=y(N+1);
D(2*N)=y(N-1)-2*y(N)+alpha*(y(N)\verb!^!2-(y(N)-y(N-1))\verb!^!2);D(N)=y(2*N);
for I=2:N-1,
D(N+I)=y(I+1)+y(I-1)-2*y(I)+alpha*((y(I+1)-y(I))\verb!^!2-(y(I)-y(I-1))\verb!^!2);
D(I)=y(N+I);
end
dy=D';
}

\end{document}